\documentclass[graybox]{svmult}

\usepackage{helvet}         
\usepackage{courier}        
\usepackage{type1cm}        
\usepackage{makeidx}         
\usepackage{graphicx}        
\usepackage{multicol}        
\usepackage[bottom]{footmisc}
\usepackage[vcentermath]{youngtab}

\usepackage{amsmath}
\usepackage{amssymb}
\usepackage[utf8]{inputenc}

\newcommand{\bea}{\begin{eqnarray}\displaystyle}
\newcommand{\eea}{\end{eqnarray}}
\newcommand{\Tr}{{\rm Tr}}

\begin{document}

\title*{Higher dimensional CFTs as 2D conformally-equivariant topological field
theories}
 \titlerunning{Higher dimensional CFTs as 2D conformally-equivariant TFTs}
\author{Robert de Mello Koch and  Sanjaye Ramgoolam}
\institute{Robert de Mello Koch \at School of Science, Huzhou University, Huzhou 313000, China and Mandelstam 
Institute for Theoretical Physics, University of Witwatersrand, Wits, 2050, South Africa, \email{robert@neo.phys.wits.ac.za}
\and Sanjaye Ramgoolam \at 
Centre for Theoretical Physics,  Department of Physics and Astronomy, Queen Mary University of London, London E1 4NS, 
United Kingdom and Mandelstam Institute for Theoretical Physics, University of Witwatersrand, Wits, 
2050, South Africa, \email{s.ramgoolam@qmul.ac.uk}\\
\\
Written version of the talk given by SR at XIV International Workshop LIE THEORY AND ITS
APPLICATIONS IN PHYSICS, Sofia}
\maketitle

\abstract{Two and three-point functions of primary fields in four dimensional CFT have a simple space-time dependences factored out from the combinatoric structure which enumerates the fields and gives their couplings. This has led to the  formulation of two dimensional topological field theories with $SO(4,2)$ equivariance which are conjectured to be equivalent to higher dimensional conformal field theories. We review this CFT4/TFT2 construction in the simplest possible setting of a free scalar field, which gives an algebraic construction of the correlators in terms of an infinite dimensional $so(4,2)$ equivariant  algebra with finite dimensional subspaces at fixed scaling dimension. Crossing symmetry of the CFT4 is related to associativity of the algebra. This construction is  then extended to describe perturbative CFT4, by making use of deformed co-products.
Motivated by the Wilson-Fisher CFT we outline the construction of U(so($d$,2)) equivariant TFT2
for non-integer $d$, in terms of diagram algebras and their representations.}

\section{Introduction}

Conformal Field Theories (CFTs) in $d>2$ dimensions have been an active topic of study in recent years. 
In part this activity has been stimulated by the AdS/CFT correspondence, originally stated as an equivalence between
${\cal N} = 4$ super Yang-Mills (SYM) theory on $R^{3,1}$ with $U(N)$ gauge group and 10 dimensional string theory\cite{Maldacena:1997re}. 
A key question is to understand how the higher dimensional quantum gravity emerges from local CFT operators and their
correlators. 
Another motivation has been to gain an understanding of exotic CFTs that do not have a conventional Lagrangian description.
Good examples of these theories include the Argyres-Douglas fixed points in 4D \cite{Argyres:1995jj} as well as the (0,2) theories in 6D \cite{Seiberg:1997zk}.
In addition, promising tools with which to study higher dimensional CFTs have become available with the revival of the 
bootstrap program\cite{Rattazzi:2008pe}, which uses associativity of the operator product expansion (OPE) to determine the CFT data.

CFTs in $d=2$ (CFT2) have been well studied since the 80’s.
The primary stimulus for this activity is the worldsheet dynamics of strings in critical string theory, described by a
CFT2 plus ghost system.
These theories have a rich structure leading to a fruitful interaction between mathematics and physics \cite{FMS,VOM}. A central role is played by 
\begin{itemize}
\item Infinite dimensional Lie algebras (the Virasoro algebra and current algebras) which control their 
spectrum and correlators.
\item The representation theory of these algebras, extended by considerations of modular transformations of characters.
\item Rational conformal field theories, with finitely many primary fields for these algebras.
\item Vertex operator algebras, which provide mathematical constructions for field operators and for the OPEs.
\end{itemize}
An important observation is that the mathematics of CFT2s use two kinds of algebras.
First, there are the symmetry algebras given by infinite dimensional Lie algebras (the Virasoro algebra, current algebra etc.).
Secondly, there is the algebra of the quantum fields themselves, formalized through vertex operator algebras.
This situation is analogous to constructions in non-commutative geometry where we have a fuzzy or quantum space which is an associative coordinate algebra \cite{Madore,Podles,NM}, as well as a Hopf algebra acting as a symmetry of 
the quantum space.
It is natural to expect a similar structure for CFT$d$, except that we have a finite dimensional symmetry algebra SO(d,2) 
replacing the Virasoro algebra (and its generalizations) and large multiplicities of irreducible representations (irreps) coming 
from the fields/quantum states.
This expectation is, at least partly, motivated by the operator/state correspondence of radial quantization
\bea
\lim_{x\to 0}{\cal O}_a(x)|0\rangle = |{\cal O}_a\rangle
\eea
which is a general property of CFTs  for any $d$.
The AdS/CFT correspondence together with the operator state correspondence implies, for example, that string states in
AdS$_5\times$S$^5$ are in correspondence with operators in ${\cal N}=4$ SYM.
An understanding of the quantum states (and associated physics) in quantum gravity on AdS spacetimes requires 
a detailed understanding of the CFT operators and associated algebraic structures.

The ${1\over 2}$-BPS sector is an interesting sector of ${\cal N} = 4$ SYM  where these ideas can be developed quite explicitly.
On the AdS side of the duality, there is a rich spectrum of physical states including gravitons, strings, branes and non-trivial
spacetime geometries.
On the CFT side of the duality, this sector is constructed from a single complex matrix $Z$, transforming in the adjoint
representation
\bea
 Z \to UZU^\dagger
\eea
under the $U(N)$ gauge symmetry.
The generic gauge invariant operator is a multi-trace operator.
For example, the complete set of gauge invariant operators that can be constructed using three fields is given by
\bea
\Tr Z^3 \qquad\qquad \Tr Z^2 \Tr Z \qquad\qquad (\Tr Z)^3
\eea
These operators have degree 3 and are in correspondence with the partitions of 3
\bea
3 = 3 \qquad\qquad 3 = 2 + 1 \qquad\qquad 3 = 1 + 1 + 1
\eea
In general, operators of degree $n$ correspond to partitions of $n$ and they can be constructed using permutations 
$\sigma\in S_n$ as follows
\bea
{\cal O}_\sigma (Z) =\sum_{i_1,\cdots,i_n} Z^{i_1}_{i_{\sigma(1)}}\cdots Z^{i_n}_{i_{\sigma(n)}}
\eea
For example
\bea
\Tr Z^3 =\sum_{i_1,i_2,i_3} Z^{i_1}_{i_2} Z^{i_2}_{i_3} Z^{i_3}_{i_1}
=\sum_{i_1,i_2,i_3} Z^{i_1}_{i_{\sigma(1)}} Z^{i_2}_{i_{\sigma(2)}} Z^{i_3}_{i_{\sigma(3)}}
\eea
with $\sigma =(123)$.
The mapping between permutations and gauge invariant operators is not one-to-one since
\bea
{\cal O}_\sigma (Z) = {\cal O}_{\gamma\sigma\gamma^{-1}}(Z)  ~~~~\hbox{ for all } ~~~ \gamma \in S_n 
\eea
which implies that two permutations in the same conjugacy class define the same gauge invariant operator.
This nicely explains why gauge invariant operators correspond to partitions of $n$.
The two-point function of degree $n$ operators is derived, as usual, by using Wick’s theorem and the basic 2-point function
\bea
\langle Z^i_j (x_1)(Z^\dagger)^k_l(x_2)\rangle = {\delta^i_l \delta^k_j\over (x_1-x_2)^2}
\eea
This allows us to express the correlator in terms of permutation group multiplications as follows \cite{Corley:2001zk}
\bea
\langle O_{\sigma_1}(Z(x_1))O_{\sigma_2}(Z^\dagger(x_2))\rangle&=&
{1\over ((x_1-x_2)^2)^n}{n!\over |{\cal C}_{p_1}|\, |C_{p_2}|}\times\cr\cr
&&\sum_{\sigma_1\in{\cal C}_{p_1}}\sum_{\sigma_2\in{\cal C}_{p_2}}\sum_{\sigma_3\in S_n}
\delta (\sigma_1\sigma_2\sigma_3)N^{C_{\sigma_3}} 
\eea
The combinatoric part of this answer is a quantity in a 2D topological field theory (TFT2).
TFT2s are equivalent to Frobenius algebras.
A Frobenius algebra is an associative algebra with a non-degenerate pairing.
The algebra corresponding to the combinatoric TFT2 of the ${1\over 2}$-BPS sector is the centre of the group algebra 
of the symmetric group $S_n$.
The connection to TFT2 can be generalized beyond the ${1\over 2}$-BPS sector and it turns out that multi-matrix sectors of 
${\cal N} = 4$ SYM are related to other Frobenius algebras, built from permutations or associated diagram algebras 
such as Brauer algebras \cite{Corley:2001zk,Balasubramanian:2004nb,deMelloKoch:2007rqf,Brown:2007xh,Bhattacharyya:2008rb,Mattioli:2014yva,Kimura:2014mka}. For a review of these ideas, the reader can consult \cite{Ramgoolam:2016ciq}.

It is natural to ask if the space-time dependence of correlators in CFT4  (and CFT$d$ for $d> 2$) can also be described
using a TFT2/Frobenius algebra language.
The paper \cite{deMelloKoch:2014aot} gives a positive answer for the case of a free 4D massless scalar field, along with 
the cases in which the scalar transforms in the fundamental or in the adjoint of a global symmetry.
The construction uses an infinite dimensional associative algebra which reproduces free field correlators of arbitrary free field
composites and is a representation of $so(4,2)$ or $Uso(4,2)$. 
This algebra has an so(4,2) invariant non-degenerate pairing.
In the paper \cite{deMelloKoch:2018klm} the algebraic structures associated with this CFT4/TFT2 construction were used 
to develop novel counting formulae and construction algorithms for the primary fields of free CFT4.
The paper \cite{deMelloKoch:2020roo} describes perturbative CFTs from this algebraic point of view (equivariant algebras).
Concrete examples that are described include sectors of $d = 4$ ${\cal N} = 4$ SYM at weak coupling as well as the 
Wilson-Fischer CFT, defined in $d = 4 -\epsilon$ using the $\phi^4$ interaction.
Novel algebraic structures needed to accomplish this include a deformed co-product for $Uso(4,2)$, the role of indecomposable
representations of $Uso(4,2)$ and diagram algebras which generalize known diagram algebras appearing in the representation theory of $Uso(d)$. 
This work has some overlap with the paper \cite{Binder:2019zqc}.

This paper is organized as follows: Section 2 reviews the CFT4/TFT2 construction in the simplest possible setting of a 
free scalar field. The result is a $U(so(4,2))$ equivariant TFT2 with the quantum field realized as a vertex operator.
Section 3 describes perturbative CFT4 by making use of deformed co-products while Section 4 introduces
diagram algebras and their representations, motivated by the Wilson-Fisher CFT.

\section{CFT4/TFT2 Construction of the free scalar field}

The axiomatic approach to TFT2 that we adopt associates geometrical objects to algebraic objects, following the standard discussions (see   the original work \cite{Atiyah} and textbooks such as  \cite{Kock}) , with appropriate adaptations to account for the infinite-dimensionality of the state spaces. 
For example, a vector space ${\cal H}$ is associated with a circle
\bea
\begin{gathered}\includegraphics[width=0.025\columnwidth]{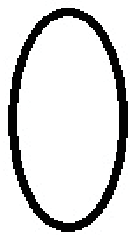}\end{gathered}\qquad\longrightarrow\qquad {\cal H}
\eea
while tensor products of ${\cal H}$ go to disjoint unions of circles
\bea
\begin{gathered}\includegraphics[width=0.025\columnwidth]{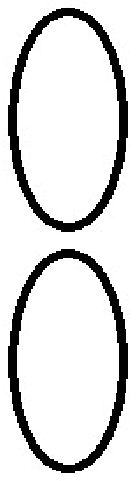}\end{gathered}
\qquad\longrightarrow\qquad {\cal H}\otimes {\cal H}
\eea
Interpolating surfaces between circles (cobordisms) are associated with linear maps between the vector spaces.
For example, the map $\delta:{\cal H}\to{\cal H}$ is represented as a cylinder
\bea
\delta_A{}^B=\quad\begin{gathered}\includegraphics[width=0.1\columnwidth]{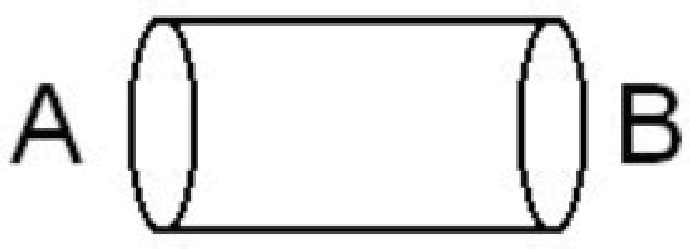}\end{gathered}
\eea
which takes circle $A$ into circle $B$, while the non-degenerate pairing $\eta:{\cal H}\otimes{\cal H}\to\mathbb{C}$
\bea
\eta_{AB}=\quad\begin{gathered}\includegraphics[width=0.05\columnwidth]{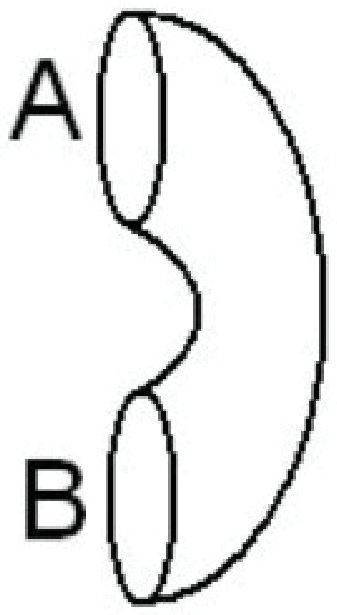}\end{gathered}
\eea
takes two circles to nothing.
The product $C:{\cal H}\otimes{\cal H}\to{\cal H}$
\bea
C_{AB}{}^D=\quad\begin{gathered}\includegraphics[width=0.1\columnwidth]{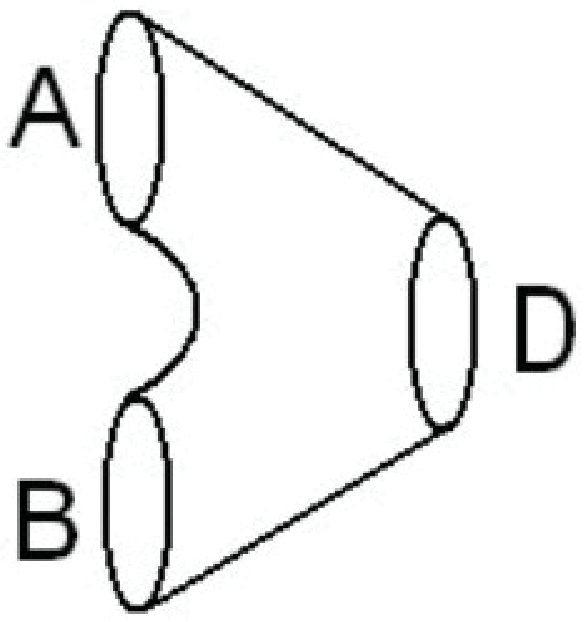}\end{gathered}
\eea
takes two circles to a circle.
In the language of category theory, the circles are objects and the interpolating surfaces (cobordisms) are morphisms in a geometrical category, while the vector spaces are objects, and the linear maps are morphisms in an algebraic category. The correspondence between geometrical objects algebraic objects is a functor between the two categories. The existence of this functor requires that all relations on the geometrical side should be mirrored on the algebraic side.
As an example, the statement that the pairing $\eta_{AB}$ is non-degenerate is expressed in terms of the inverse pairing
\bea
\tilde\eta^{AB}=\quad\begin{gathered}\includegraphics[width=0.065\columnwidth]{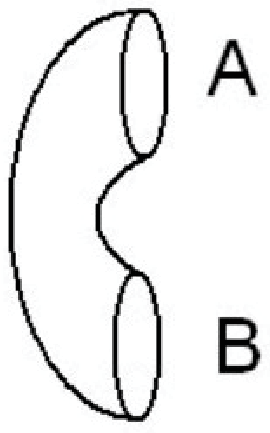}\end{gathered}
\eea
as the statement that $\eta$ and $\tilde\eta$ glue to give the cylinder
\bea
\eta_{AB}\tilde\eta^{BC}=\delta_A{}^C
\qquad\qquad\begin{gathered}\includegraphics[width=0.225\columnwidth]{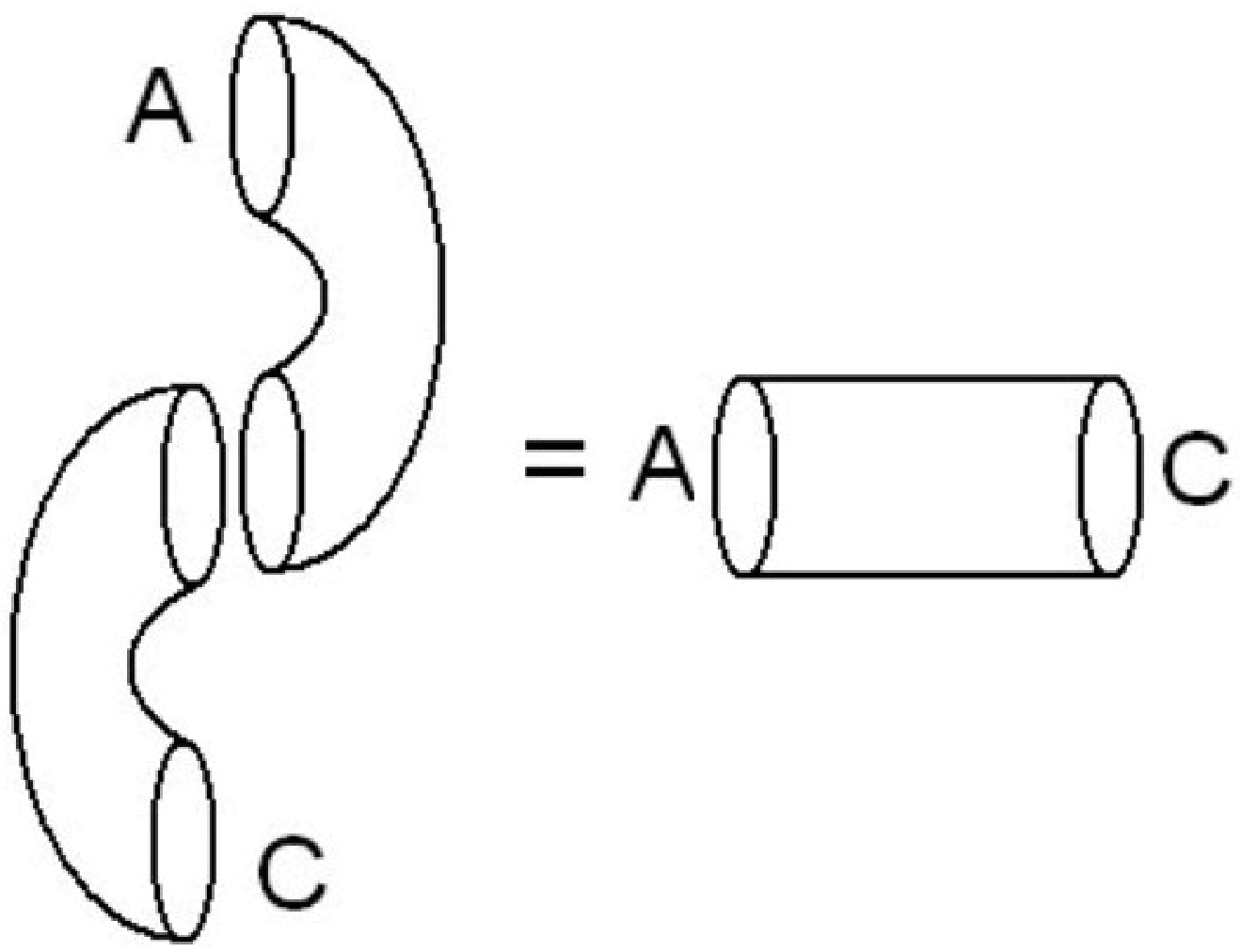}\end{gathered}
\label{snakecylinder}
\eea
where the gluing operation is implemented by summing over the circles to be glued.
Using the product $C_{AB}{}^D$ and the pairing $\eta_{AB}$ we can define a new map 
\bea
C_{ABD}=\eta_{DC}C_{AB}{}^D
\qquad\qquad\begin{gathered}\includegraphics[width=0.225\columnwidth]{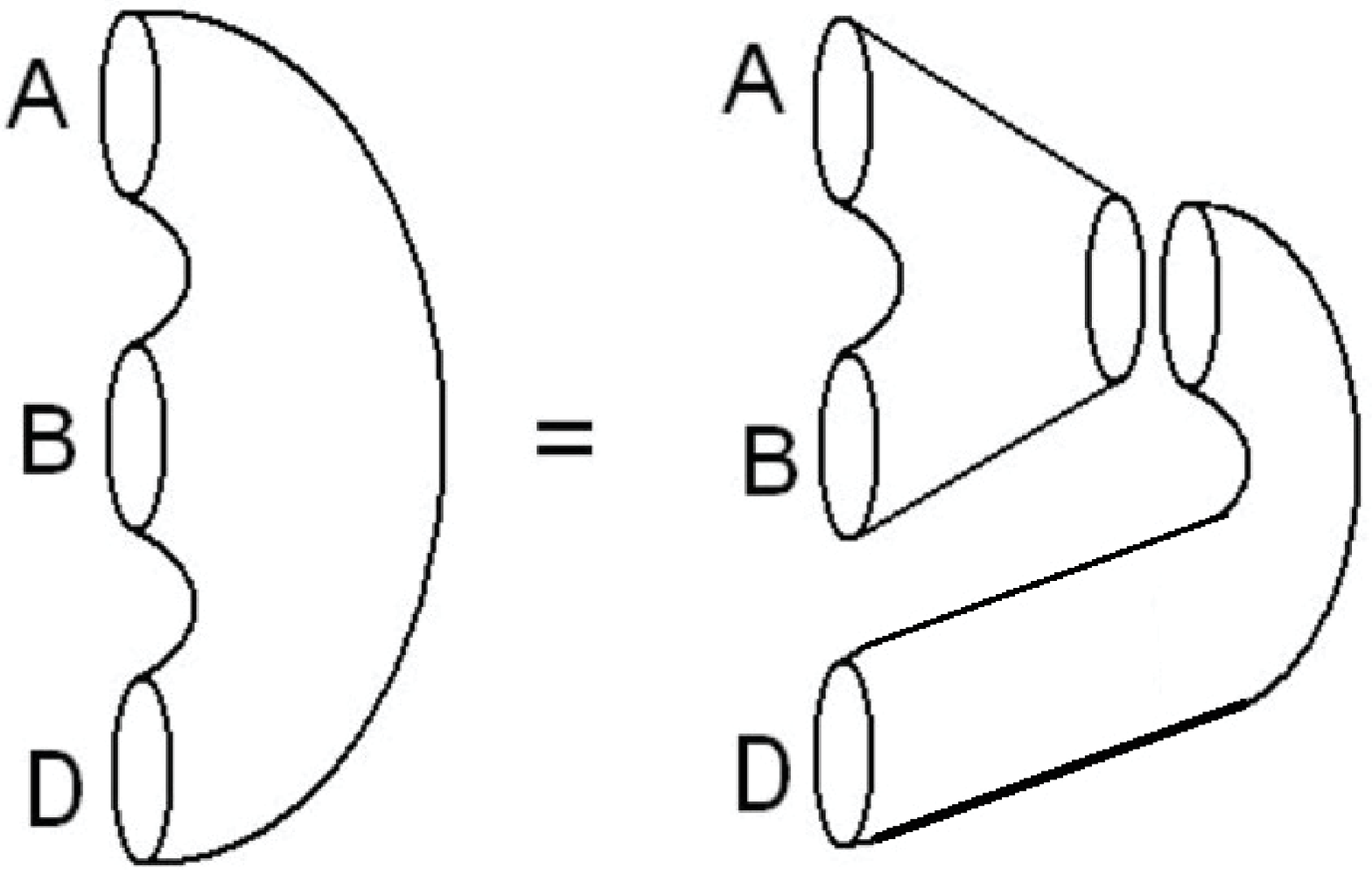}\end{gathered}
\label{opecorr}
\eea
which is the familiar relation between the CFT correlator ($C_{ABD}$) and the OPE ($C_{AB}{}^D$).
Finally, associativity of the OPE is expressed as
\bea
C_{AB}{}^EC_{EC}{}^D=C_{BC}{}^EC_{EA}{}^D
\qquad\qquad\begin{gathered}\includegraphics[width=0.35\columnwidth]{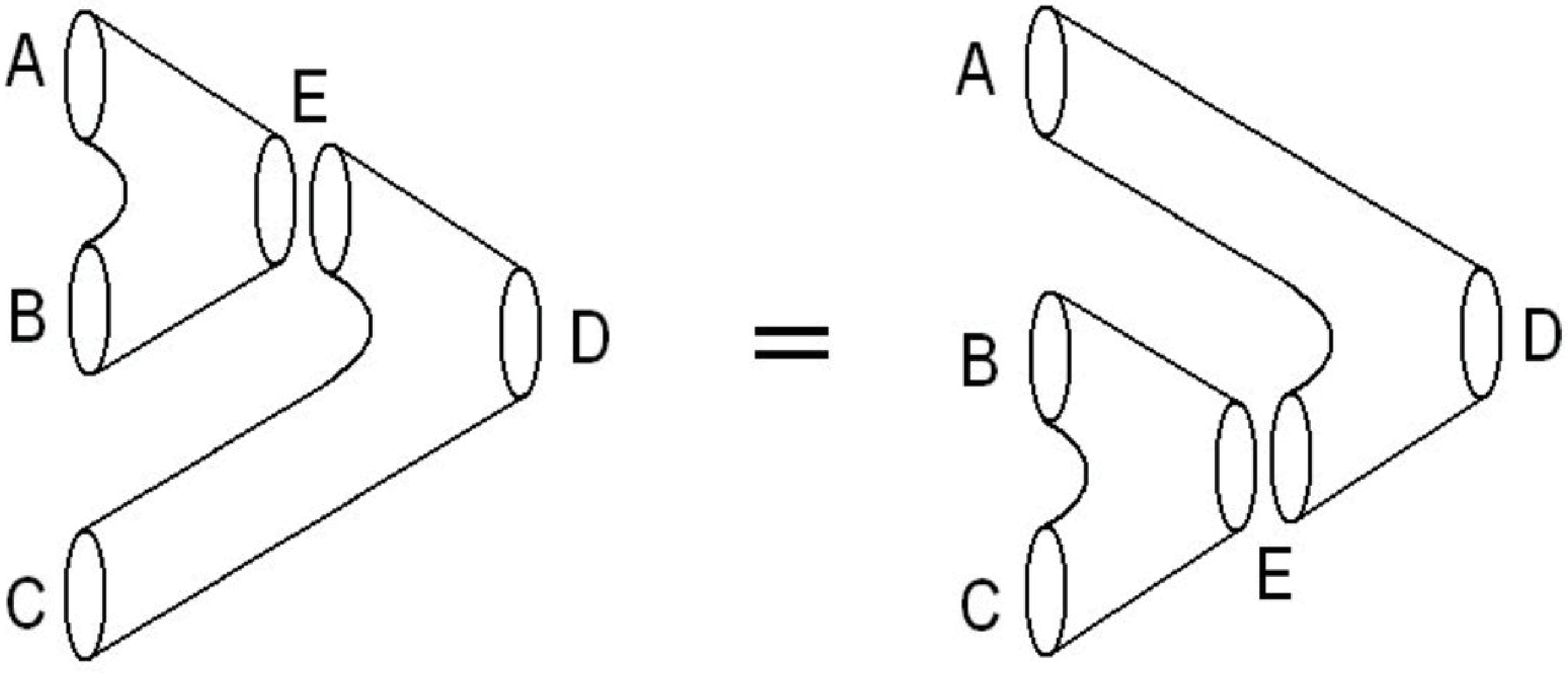}\end{gathered}
\eea
To summarise, TFT2’s correspond to commutative, associative, non-degenerate algebras known as Frobenius algebras.
TFT2 with a global symmetry group $G$ is defined by \cite{Moore:2006dw}. Since this will play an important role in our construction, it is worth summarizing the essential features from \cite{Moore:2006dw} with one important modification of the discussion due to the infinite dimensionality of the state space 
\begin{itemize}
\item[1.] The state space is a representation of a group $G$ - which will be $SO(4,2)$ in our application.
\item[2.] The linear maps are $G$-equivariant linear maps.
\item[3.] The state space is infinite dimensional: amplitudes are defined for surfaces without handles. Consequently, this is a genus restricted TFT2.
\end{itemize}

The basic two point function in the CFT of a free scalar in four dimensions is
\bea
\langle\phi(x_1)\phi(x_2)\rangle ={1\over (x_1-x_2)^2}
\eea
Correlators of composite operators are constructed using this contraction, according to Wick's theorem.
Thus, the first step in our construction is to understand the 2-point function of the elementary field in the TFT2 language.
The $so(4,2)$ symmetry of the CFT is our starting point. 
The Lie algebra is spanned by the dilatation opertor $D$ which generates dilatations, the momenta $P_\mu$ which 
generate translations, the generators $M_{\mu\nu}$ of $so(4)$ rotations and the generators $K_{\mu}$ of special 
conformal transformations. 
To carry out a radial quantization of the theory we choose a point, say the origin of Euclidean $\mathbb{R}^4$. The usual state operator map associates states with the scalar field and its descendents
\bea
\lim_{x\to 0}\phi(x)|0\rangle &=& v^+\cr
\lim_{x\to 0}\partial_\mu\phi(x)|0\rangle &=& P_\mu v^+\cr
&\vdots&
\eea
The state $v^+$ is the lowest energy state in a lowest-weight representation $V_+$ of $so(4,2)$
\bea
Dv^+ &=& v^+\cr
K_\mu v^+ &=& 0\cr
M_{\mu\nu}v^+ &=& 0
\eea
Higher energy states are generated by $S_l^{\mu_1\mu_2\cdots\mu_l}P_{\mu_1}P_{\mu_2}\cdots P_{\mu_l}v^+$,
where $S_l^{\mu_1\mu_2\cdots\mu_l}$ is a symmetric traceless tensor of $so(4)$. The index $l$ labels a basis of linearly independent symmetric traceless tensors. There is also a dual representation $V_-$, which is a representation with negative scaling dimensions
\bea
Dv^- &=& -v^-\cr
P_\mu v^- &=& 0\cr
M_{\mu\nu}v^- &=& 0
\eea
Other states in this representation are generated by acting with $S_l^{\mu_1\mu_2\cdots\mu_l}K_{\mu_1}$ 
$\cdots K_{\mu_l}$.
There is a $\eta: V_+\otimes V_-\to \mathbb{C}$, which is so(4,2) invariant
\bea
\eta ({\cal L}_av,w) + \eta(v,{\cal L}_aw) = 0\label{InvCond}
\eea
After making the choice
\bea
\eta (v^+,v^-) = 1
\eea
the invariance conditions (\ref{InvCond}) and the properties of the states $v^+$ and $v^-$, determine $\eta$. For example
\bea
\eta (P_\mu v^+, K_\nu v^-) &=& -\eta (v+,P_\mu K_\nu v^-)\cr
&=& \eta(v^+,(-2D\delta_{\mu\nu}+2M_{\mu\nu})v^-) = 2\delta_{\mu\nu}
\eea
Using invariance conditions one finds that $\eta(P_\mu P_\mu v^+,v^-)$ is zero. Setting $P_\mu P_\mu v^+=0$, which physically corresponds to imposing the equation of motion, identifies $V_+$ as a quotient of a bigger representation $\tilde{V}_+$. $\tilde{V}_+$ is spanned by
\bea
P_{\mu_1}\cdots P_{\mu_l} v^+
\eea
i.e it is the vector space of polynomials in $P_\mu$. This is an indecomposable representation. After we perform the quotient by the equation of motion, we recover the irreducible representation $V_+$. The quotient also ensures that $\eta$ is non-degenerate i.e. that there are no null vectors. So we see that $\eta$ is the structure we need for the construction of a TFT2 with so(4,2)
symmetry. It has both the non-degeneracy property and the invariance property.
So there is an invariant in $V_+\otimes V_-$ and thus in $V_-\otimes V_+$, but not in $V_+\otimes V_+$ or $V_-\otimes V_-$. It is useful to introduce $V=V_+\oplus V_-$ and define $\hat\eta:V\otimes V\to\mathbb{C}$ 
\bea
\hat\eta =\left(\begin{array}{cc} 0 &\eta_{+-}\\ \eta_{-+} &0\end{array}\right)
\eea
In $V$ we have a state, corresponding to the quantum field, given by
\bea
\Phi (x)={1\over \sqrt{2}}(e^{-iP\cdot x}v^+ + x^{\prime 2}e^{iK \cdot x'}v^-)\qquad 
x'_\mu = {x_\mu\over x^2}
\eea
and a calculation with the invariant pairing shows that
\bea
\eta(\Phi(x_1),\Phi(x_2)) = {1\over (x_1-x_2)^2}
\eea
This is the basic free field 2-point function, now constructed from the invariant map $\eta : V\otimes  V\to \mathbb{C}$.
To get all correlators, we must set up a state space, which knows about composite operators.
The states obtained by the standard operator state correspondence from general local operators are of the form
\bea
P_{\mu_1}\cdots P_{\mu_{n_1}}\phi P_{\nu_1}\cdots P_{\nu_{n_2}}\phi\cdots P_{\tau_1}\cdots P_{\tau_{n_m}}\phi
\eea
Composite operators belonging to the $n$ field sector correspond to states in which $n$ $\phi$ fields appears.
Particular linear combinations of these states are primary fields, which are lowest weight states (annihilated by $K_\mu$) 
that generate irreducible representations (irreps) of SO(4,2) through the action of the raising operators ($P_\mu$).
The list of primary fields in the $n$-field sector is obtained by decomposing the space
\bea
{\rm Proj}_{\hbox{$S_n$ invt} }(V^{\otimes n}_+) \equiv {\rm  Sym}^n (V_+)\label{CompPrim}
\eea
into SO(4,2) irreps.
The symmetrization on the right hand side of (\ref{CompPrim}) is needed because $\phi$ is a boson.
We can now introduce the state space ${\cal H}$ of the TFT2, which we associate to a circle in TFT2.
The state space consists of all possible primaries and their descendents
\bea
{\cal H} =\bigoplus_{n=0}^\infty {\rm Sym}^n(V)\label{frstss}
\eea
where $V = V_+ \oplus V_-$.
This state space is big enough to accommodate all the composite operators and it admits an invariant pairing.
The state space is small enough for the invariant pairing to be non-degenerate
The state space contains
\bea
\Phi (x)\otimes \Phi (x)\otimes\Phi \cdots\otimes\Phi (x)
\eea
which is used to construct composite operators in the TFT2 set-up.
By construction, the pairing $\eta :{\cal H}\otimes {\cal H}\to \mathbb{C}$ reproduces all 2-point functions of arbitrary
composite operators. 
The construction is straight forward: recall ${\cal H}$ is built from tensor products of $V$, and we have already introduced
an ``elementary'' $\hat\eta:V\otimes V\to\mathbb{C}$. 
The construction of the complete $\eta$ map is built from products of the elementary $\hat\eta$, using Wick contraction sums,
in the obvious way. As an example, for $v_1,v_2,v_3.v_4\in V$ we have
\bea
\eta (v_1\otimes v_2,v_3\otimes v_4)=\hat{\eta}(v_1,v_3)\hat{\eta}(v_2,v_4)+\hat{\eta}(v_1,v_4)\hat{\eta}(v_2,v_3)
\eea
We complete the definition by setting
\bea
\eta (v^{(n)},v^{(m)})\propto\delta^{mn}
\eea
where $v^{(k)}\in {\rm Sym}^{ k } (V)$.
This defines the pairing $\eta_{AB}$ where $A,B$ take values in the space ${\cal H}$ given by the sum of all $n$-fold 
symmetric products of $V=V_+\oplus V_-$. 
Notice that the building blocks used in constructing $\eta$ are invariant maps.
The product of these invariant maps is also obviously invariant. 
We can also demonstrate that $\eta$ is non-degenerate. 
The basic idea is that if you have a non-degenerate pairing $V\otimes  V\to\mathbb{C}$, it extends to a
non-degenerate pairing on ${\cal H}\otimes {\cal H}\to \mathbb{C}$, by using the sum over Wick patterns. 
Consequently, we have
\bea
\eta_{AB}\tilde{\eta}^{BC} = \delta_A^C
\eea
This is the snake-cylinder equation, given in (\ref{snakecylinder}).

In a very similar way it is possible to define 3-point functions $C_{ABC}$ and, in general higher point functions $C_{ABC\cdots}$
using Wick pattern products of the basic $\eta$’s. By writing explicit formulae for these sums over Wick patterns, we can show that the associativity equations are satisfied.
Consider equation (\ref{opecorr}).
The $C_{ABC}$ give 3-point functions, while the $C^C_{AB} = C_{ABD}\tilde\eta^{DC}$ give the OPE-coefficients. 
Through this connection, the associativity equations of the TFT2 are the crossing equations of CFT4, obtained by 
equating expressions for a 4-point correlator obtained by doing OPEs in two different ways. 
There an important property of the OPE in this language, easily illustrated by the product
\bea
{\rm Sym}^2(V)\otimes {\rm Sym}^2(V) \to {\rm Sym}^4(V) \oplus {\rm Sym}^2(V) \oplus \mathbb{C}
\eea
which corresponds to the free field OPE, which takes the schematic form
\bea
\phi^2(x)\phi^2(0) \to \phi^4 \oplus \phi^2 \oplus 1
\eea
This demonstrates that the presence of both $V_+$ and $V_-$ is needed if the TFT2 is to construct this OPE in 
representation theory.

The algebraic framework developed above allows us to exhibit novel ring structures in the state space of the TFT2.
Further, this algebraic structure can profitably be used to give a construction of primary fields in free CFT4/CFT$d$
\cite{deMelloKoch:2018klm}. 
The state space in radial quantization, set up around $x=0$, is (for $x' =0$, we would keep $V_-$ instead)
\bea
{\cal H}_+ =\bigoplus_{n=0}^\infty {\rm Sym}^n(V_+)
\eea
The irrep $V_+$ is isomorphic to a space of polynomials in variables $x_\mu$, quotiented by the ideal generated by $x_\mu x_\mu$.
Taking a many-body physics view of ${\cal H}_+$, this is a quotient of a polynomial ring in $dn$ variables $x^I_\mu$.
The construction of primaries, or equivalently, the problem of describing lowest weight states of irreducible representations in 
${\cal H}_+$ is usefully done by recognizing the connection to a closely related problem about rings.
It turns out that the construction of primary fields in $d$ dimensions, and the refined counting of these primaries, according 
to their scaling dimension $n$ and $so(d)$ irreps, is equivalent to studying a polynomial ring in variables the $X^A_\mu$ with
$\mu\in\{1,\cdots,d\}$ and $A\in\{1,2,\cdots,n-1\}$, under the constraints
\bea
A(1-A^2)\sum_{\mu=1}^d X^A_{\mu}X^A_{\mu}&+&\sum_{B:B>A}\sum_{\mu=1}^d 2A(1 + A)X^A_\mu X^B_\mu\cr\cr 
&+&\sum_{B:B<A}\sum_{\mu=1}^d B(1 + B) X^B_\mu X^B_{\mu} = 0
\eea
for $1\le A\le (n-1)$, and
\bea
\sum_{A=1}^{n-1}\sum_{\mu=1}^d X^A_\mu X^A_\mu = 0
\eea
The details of the derivation of these constraints are explained in  \cite{deMelloKoch:2018klm}. 

\section{Perturbative CFTs}

Having explained the CFT4/TFT2 construction for the free scalar field, it is natural to ask about constructions for interacting
theories. Towards this end, the first example theory we have in mind is the Wilson-Fischer (WF) fixed point, described by the
Lagrangian
\bea
\int d^d x(\partial_\mu\phi \partial^\mu\phi +{g\over 4!}\phi^4)
\eea
together with a continuation of the Feynman rules to $d = 4-\epsilon$ dimensions. 
Choosing the critical value of the coupling constant
\bea
g^*\sim {16\pi^2\over 3}\epsilon + O(\epsilon^2)
\eea
leads to a vanishing beta function and, consequently, a CFT.
The fundamental field $\phi$ as well as composite operators (given by polynomials in derivatives of $\phi$) have a 
modified dimension. 
Apart from the classical dimension, there is also an anomalous dimension, generated by loop corrections. 
The anomalous dimensions of the WF theory are captured by a dilatation operator. 
In particular, the one-loop corrections to the dimensions of composite operators
\bea
\partial^{k_1}\phi\partial^{k_2}\phi \cdots \partial^{k_L}\phi
\eea
are captured by a 2-body Hamiltonian
\bea
H = \sum_{i<j}\rho_{ij} (P_0)
\eea
where $P_0$ is a projector to an irrep in $V\otimes V$ with $V$ the irrep of the scalar $\phi$\cite{Liendo:2017wsn}.
At order $\epsilon$, $\phi$ has a vanishing anomalous dimension, while that of $\phi^2$ is non-vanishing.
A naive intuition informed by tensor products of representations would suggest that dimension of a composite operator
is given by the sum of the dimensions of its constituents, but this is not correct.
To obtain the correct dimension for $\phi^2$ we need
\bea
D(v \otimes v) = (D\otimes  1+1\otimes D)(v\otimes  v) +{\epsilon\over 3} P_0(v\otimes v)
\eea
This motivates the definition of the deformed co-product
\bea
\Delta (D) = D\otimes 1 + 1\otimes D + {\epsilon\over 3}P_0
\eea
This deformation is highly reminiscent of deformations we encounter in quantum groups.
For example, in $U_q(su(2))$ we have
\bea
\Delta(J_+) = J_+\otimes q^H + q^{-H}\otimes  J_+
\eea
and
\bea
\Delta ({\cal L}_a) \in {\rm End}(V\otimes W)\qquad  \Delta_\epsilon({\cal L}_a) \in {\rm End}(V\otimes W)
\eea
with
\bea
\Delta ({\cal L}_a) &=& \Delta_0({\cal L}_a) + \epsilon \Delta_\epsilon ({\cal L}_a)\cr
\Delta_0({\cal L}_a) &=& {\cal L}_a\otimes 1 + 1\otimes{\cal L}_a
\eea
such that
\bea
[{\cal L}_a,{\cal L}_b]&=&f^c_{ab}{\cal L}_c\cr
[\Delta({\cal L}_a), \Delta({\cal L}_b)] &=& f^c_{ab}\Delta({\cal L}_c)
\eea
At order $\epsilon$ we have worked out the deformation needed to explain the complete spectrum of one loop
anomalous dimensions.
The co-products for the complete set of generators are
\bea
\Delta(D) &=& D \otimes 1 + 1\otimes D + {\epsilon\over 3} P_0\cr
\Delta(P_\mu) &=& P_\mu\otimes 1 + 1\otimes  P_\mu\cr
\Delta (K_\mu) &=& K_\mu\otimes 1 + 1\otimes  K_\mu - {\epsilon\over 3}P_0
\left( {\partial\over\partial P_\mu}\otimes 1 + 1 \otimes {\partial\over\partial P_\mu}\right) P_0\cr
\Delta(M_{\mu\nu}) &=& M_{\mu\nu}\otimes 1 + 1\otimes M_{\mu\nu}
\eea
It is a straightforward exercise to verify that the above co-products are consistent with the commutation relations of $so(4,2)$. For example, we have checked that
\bea
[\Delta (K_\mu), \Delta(P_\nu)] = 2\Delta (M_{\mu\nu}) - 2\delta_{\mu\nu}\Delta(D)
\eea
In performing this check and others like it, it is useful to note $\Delta_0({\cal L}_a)P_0 = P_0\Delta_0({\cal L}_a)$ 
and $P^2_0 = P_0$.

The planar $SU(2)$  sector in ${\cal N} = 4$ SYM is another interacting CFT that has an instructive TFT2/CFT4 construction.
For this example there is no need to continue to $d=4-\epsilon$ dimensions. 
In this example too, deformed co-products are needed to reproduce the one loop spectrum of anomalous dimensions. 
Consider the three operators
\bea
{\cal O}_z &=&{1\over\sqrt{2}N}{\rm Tr}(Z^2)\qquad 
{\cal O}_y\,\,\,=\,\,\,{1\over \sqrt{2}N}{\rm Tr}(Y^2)\cr
{\cal O}_{zy}&=&{1\over\sqrt{3}N^2}\left({\rm Tr}(YZYZ)-{\rm Tr}(Y^2Z^2)\right)
\eea
These operators are all eigenstates of the one loop dilatation operator. 
${\cal O}_{zy}$ has a non-zero anomalous dimension $\delta = {3\lambda\over 4\pi^2}$ in the planar 
limit\cite{Beisert:2003tq}.
The anomalous dimensions of both ${\cal O}_z$ and ${\cal O}_y$ vanish.
In the free theory, the dimensions add
\bea
{\rm Dim}({\cal O}_z )+{\rm Dim}({\cal O}_y)={\rm Dim}({\cal O}_{zy})
\eea
At first order in the interaction this relationship is corrected as follows
\bea
{\rm Dim}({\cal O}_z) +{\rm Dim}({\cal O}_y ) = {\rm Dim}({\cal O}_{zy} ) - \delta
\eea
As the first step, consider so(4,2) irrep generated by ${\cal O}_z$.
In the operator-state correspondence, the operator ${\cal O}_z$ corresponds to a tower of operators
\bea
{\cal O}_z (0) &\to& v_z\cr
\partial_{\mu_1}{\cal O}_z (0) &\to& P_{\mu_1}v_z\cr
\partial_{\mu_1}\partial_{\mu_2}{\cal O}_z (0) &\to& P_{\mu_1}P_{\mu_2}v_z\cr
&\vdots&
\eea
The states live in a representation $V_z$ of $so(4,2)$. The lowest weight state $v_z$ has the properties
\bea
Dv_z = 2v_z \qquad M_{\mu\nu} v_z = 0 \qquad K_\mu v_z = 0
\eea
At dimension $(2 + k)$ we have states
\bea
V_k = {\rm Span}\left\{ P_{\mu_1}\cdots P_{\mu_k} v_z\right\}
\eea
The direct sum forms the $so(4; 2)$ irrep $V^{(z)}$
\bea
V =\bigoplus_{k=0}^\infty V_k
\eea
There is a similar representation $V_y$ built on the primary ${\cal O}_y$.
$V_z$ and $V_y$ are isomorphic representations of so(4,2).
We also need the representation $V_{zy}$, built on ${\cal O}_{zy}$. 
This representation has lowest weight state $v_{zy}$ with properties
\bea
Dv_{zy} &=& (4 +\delta)v_{zy}\cr
K_{\mu}v_{zy} &=& 0\cr
M_{\mu\nu}v_{zy} &=& 0
\eea
States at $D = 4 + \delta + k$ are obtained by acting with $k$ $P$’s.

Given the non-additivity of the anomalous dimensions, we cannot model the 3-point correlator with the standard action 
of the Lie algebra on $V_2\otimes V_2$. If we use the standard action, we would have
\bea
\Delta_0(D)(v_z\otimes v_y) = (D\otimes 1 + 1\otimes D)(v_z\otimes v_y ) = 4v_z\otimes v_y
\eea
whereas the dimension of $v_{zy}$ is $4+\delta$. The map $f:v_{zy}\to v_z\otimes v_y$
\bea
\Delta_0(D)f (v_{zy}) = f\Delta_0(D)(v_{zy})
\eea
can be extended to an equivariant map $V_{zy}\to V_z\otimes V_y$ at zero coupling, but cannot be so extended 
when we turn on $\delta$ at non-zero coupling.

Let $P_4$ be the projector to $V_4$ - the so(4,2) representation with scalar lowest weight of dimension 4 - in the standard 
tensor product decomposition of $V_2\otimes V_2$. We can define a deformed co-product
\bea
\Delta(D) &=& \Delta_0(D) + \delta P_4\cr
\Delta (P_\mu) &=& \Delta_0(P_\mu)\cr
\Delta(M_{\mu\nu}) &=& \Delta_0(M_{\mu\nu})\cr
\Delta (K_{\mu}) &=& \Delta_0(K_\mu) - {\delta\over 2}P_4\Delta_0\left({\partial\over\partial P_\mu}\right)P_4
\eea
With the so(4,2) action on $V_z\otimes V_y$ given by
\bea
{\cal L}_a : v_1\otimes v_2 \to \Delta({\cal L}_a)(v_1\otimes v_2)
\eea
and the so(4,2) action on $V_{zy}$ which we will refer to as $\rho_{zy}$, we can extend $f$
\bea
f : V_{zy} \to V_z\otimes V_y
\eea
such that
\bea
f \rho_{zy} ({\cal L}_a) = \Delta({\cal L}_a)f
\eea
Using the map $f$, we can construct the correlator as follows
\bea
\eta((e^{-iP\cdot x_1}v^+\otimes e^{-iP\cdot x_2}v^+), (x'_3)^2 f(e^{iK\cdot x'_3} v^+_{zy}))
\eea
The inner product $g$ on $V_z\otimes V_y$ is related by using the anti-automorphism on $so(4,2)$ to the invariant pairing on
\bea
\eta : (V_+\otimes V_+)\otimes (V_-\otimes V_-) \to \mathbb{C}
\eea

\section{$d = 4 -\epsilon$ and diagram algebras}

In our example of the Wilson-Fischer, we need to continue from $d=4$ to $d=4-\epsilon$ dimensions in order to obtain
a non-trivial CFT. 
Analytically continued tensor rules, and in particular the rule $\delta_\mu^\mu = 4 - \epsilon$,
are needed to construct the stress tensor with the right properties.
The state space $V_+$ used in the free scalar field theory is a quotient of a space $ \tilde V_+ $ spanned by states of the form  
\bea
\left\{ P_{\mu_1}\cdots P_{\mu_k} v\right\}
\eea
The quotient amounts to setting to zero $P_{\mu} P_{\mu} v $.  The stress tensor
\bea
T_{\mu\nu} &=& {1\over 2} (P_\mu v\otimes P_\nu v + P_\nu v\otimes P_\mu v 
-\delta_{\mu\nu} P_\tau v\otimes P_\tau v)\cr
&&-{\alpha\over 6} \Delta (P_\mu P_\nu - P^2\delta_{\mu\nu})v\otimes v
\eea
is a state in $\tilde{V}_+\otimes\tilde{V}_+ $.
The above state is conserved and traceless upon using the interacting equation of motion, along with
\bea
Dv=\left(1-{\epsilon\over 2}\right)v\qquad M_{\alpha\beta}v=0\qquad \delta_{\mu\mu} = 4 -\epsilon
\eea
The positive part of the state space
\bea
\bigoplus_{n=0}^\infty {\rm Sym}^n (V_+)
\eea
where $V_+ = \tilde{V}_+/\{P^2v\}$ is replaced by
\bea
\bigoplus_{n=0}^\infty {\rm Sym}^n (\tilde{V}_+)
\eea
and we need to quotient by
\bea
P^2 v - 4g^* v\otimes v\otimes v
\eea
Thus, understanding the interacting equations of motion in terms a quotient space, requires working  with  $\tilde{V}_+$ and its tensor powers. The quotient condition relates states in $ \tilde V_+$ to $ \tilde V_+^{ \otimes 3}$. 
Notice that in both the free and interacting theories, we move from $V_+$ to $\tilde{V}_+$ by quotienting with 
the equation of motion.

To make sense of the rule $\delta_{\mu\mu} = 4 -\epsilon$, we need to construct a diagram algebra, much like the Brauer algebras. In our TFT2 setting, $Uso(d)$ (and $Uso(d,2)$) itself has to be made diagrammatic in order to give a TFT2 
with conformal equivariance formulation of the perturbative correlators.

If we depict the product $M_{ij}M_{kl}$ in the universal enveloping algebra Uso($d$) by juxtaposing two boxes side to side, 
we can express
\bea
M_{ij}M_{kl}-M_{kl}M_{ij}=\delta_{jk}M_{il}+\delta_{il}M_{jk}-\delta_{jl}M_{ik}-\delta_{ik}M_{jl}
\eea
as a relation between diagrams as follows
\bea
&&\begin{gathered}\includegraphics[scale=0.4]{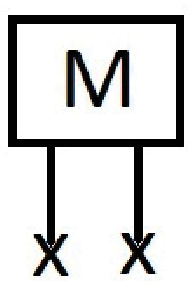}\end{gathered}
\begin{gathered}\includegraphics[scale=0.4]{M}\end{gathered}
-\begin{gathered}\includegraphics[scale=0.4]{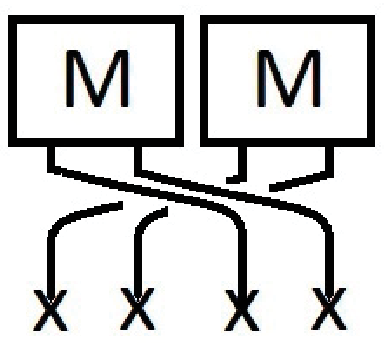}\end{gathered}=
\begin{gathered}\includegraphics[scale=0.4]{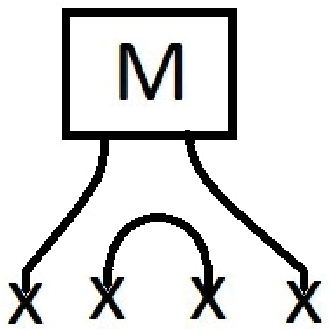}\end{gathered}
+\begin{gathered}\includegraphics[scale=0.4]{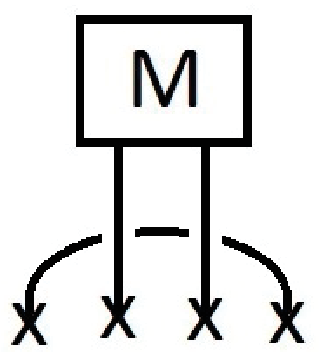}\end{gathered}
-\begin{gathered}\includegraphics[scale=0.4]{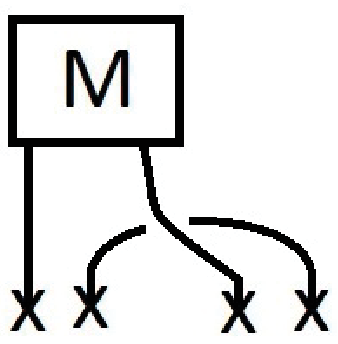}\end{gathered}
-\begin{gathered}\includegraphics[scale=0.4]{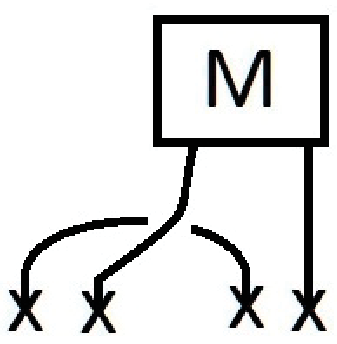}\end{gathered}\cr
&&\label{fig:Mcomm}
\eea
To go from the diagrammatic relation to the equation in Uso($d$), we attach the labels $i,j,k,l$ to the crosses starting
with $i$ for the left-most cross and proceeding with $j,k,l$ as we go to the crosses towards the right.
The antisymmetry can be expressed diagrammatically as follows
\bea
\begin{gathered}\includegraphics[scale=0.4]{M}\end{gathered}=
-\begin{gathered}\includegraphics[scale=0.4]{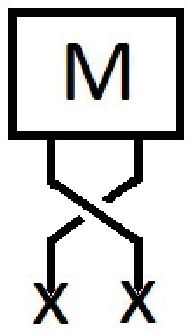}\end{gathered}\label{fig:antisymmM}
\eea
The quadratic Casimir  $ M_{ij} M_{ ij}$ is associated to the diagram shown below
\bea
\begin{gathered}\includegraphics[scale=0.4]{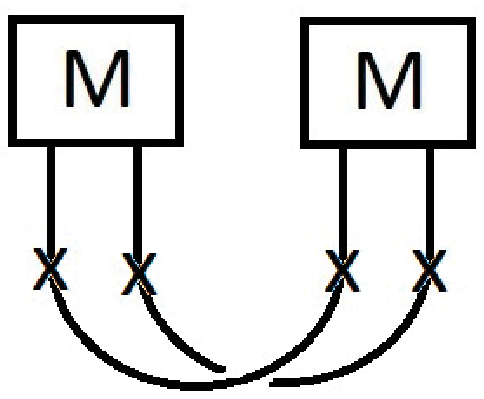}\end{gathered}=
\begin{gathered}\includegraphics[scale=0.4]{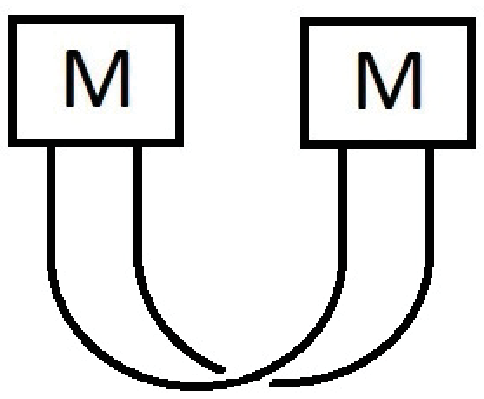}\end{gathered}\label{CasDiag}
\eea

We will define an infinite dimensional associative algebra over $\mathbb{C}$, denoted ${\cal F}$, abstracted from the 
generators $M_{ij}$ of Uso($d$). An associative algebra is a vector space equipped with a product $m$
\bea
m :{\cal F}\otimes {\cal F} \to {\cal F}
\eea
The vector space ${\cal F}$ is
\bea
{\cal F} = \mathbb{C} \oplus {\rm Span}_{\mathbb{C}}(M) \oplus\cdots
\eea
The $\cdots$ refers to subspaces which can be specified efficiently, using the oscillator construction of $Uso(d)$ and its
interpretation in terms of equivariant maps and diagrams.
The $d$-dimensional oscillator relations are
\bea
[a^\dagger_i , a_j ] = -\delta_{ij}
\eea
and the Lie algebra generators of so($d$) can be written as
\bea
M_{ij} = a^\dagger_i a_j - a^\dagger_j a_i
\eea
Think of this as specifying a state (which we can also call $M_{ij}$) using $V = {\rm Span} (a,a^\dagger)$ and
$W = {\rm Span} (e_i : i\in 1,\cdots, d)$
\bea
M_{ij}=a^\dagger\otimes e_i\otimes a\otimes e_j-a^\dagger\otimes e_j\otimes a\otimes e_i \in V\otimes W\otimes V\otimes W
\eea
It is useful to write this as
\bea
M_{ij} = P^{W\otimes W}_A (a^\dagger\otimes e_i\otimes a\otimes e_j )
\eea
The number of these $M_{ij}$ is $d(d-1)/2$. 
This obstructs continuing $d$ to non-integer dimensions.
In contrast to this, the space of equivariant maps
\bea
P_A(W\otimes W) \to P^{W\otimes W}_A (V_+\otimes W\otimes V_-\otimes W)
\eea
is a one-dimensional vector space (for $d > 4$\footnote{For $d=4$ we can also use $\epsilon_{i_1i_2i_3i_4}$ which 
gives another map, so we will use large $d$ in the appropriate places in our definitions to keep things as simple as possible.})
spanned by the map $M$ acting as
\bea
&&M : (e_{i_1}\otimes e_{i_2}-e_{i_2}\otimes e_{i_1}) 
\to (a^\dagger\otimes e_{i_1}\otimes a\otimes e_{i_2}-a^\dagger\otimes
e_{i_2}\otimes a\otimes e_{i_1})\cr
&&
\eea
More compactly, we can write
\bea
M : W_2 \to (VW)_2
\eea
where $W_2 = P_A(W\otimes W)$ and
\bea
(VW)_2 = P^{W\otimes W}_A (V_+\otimes W\otimes V_-\otimes W)
\eea
Now, introduce the infinite dimensional associative algebra
\bea
{\cal F} =\oplus_{m,n=0}{\cal F}_{m,n}
\eea
with
\bea
{\cal F}_{m,n} = {\rm Hom}_{so(d):d>2m+2n}(W^{\otimes m}_2 , (VW)^{\otimes n}_2)
\eea
This algebra contains all the $M$-diagrams we drew earlier, and it includes diagram with arbitrarily large numbers of $M$-boxes.
We define $U_*$ as a quotient of this space of diagrammatic maps by the commutation relation (\ref{fig:Mcomm}).

In order to understand the representation theory of $U_*$, which will be a diagrammatic analog of the representation theory 
of Uso($d$) at large $d$, we will start by interpreting the basic equation
\bea
[M_{ij},a^\dagger_k] = \delta_{jk}a^\dagger_k - \delta_{ik}a^\dagger_j
\eea
which gives the action of $Uso(d)$ on the $d$-dimensional vector representation. By using labelled $M$-box diagrams, 
and associating to $a^\dagger_k$ a line joining a cross to a circle, the above equation becomes
\bea
\begin{gathered}\includegraphics[scale=0.4]{M}\end{gathered}\,\,\,
\begin{gathered}\includegraphics[scale=0.4]{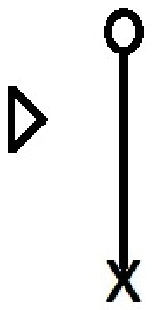}\end{gathered}
&=&\begin{gathered}\includegraphics[scale=0.4]{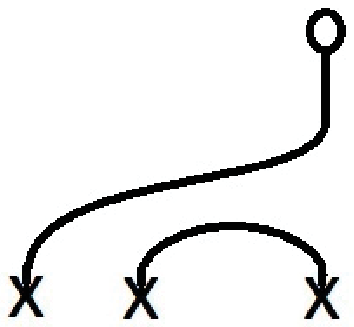}\end{gathered}
-\begin{gathered}\includegraphics[scale=0.4]{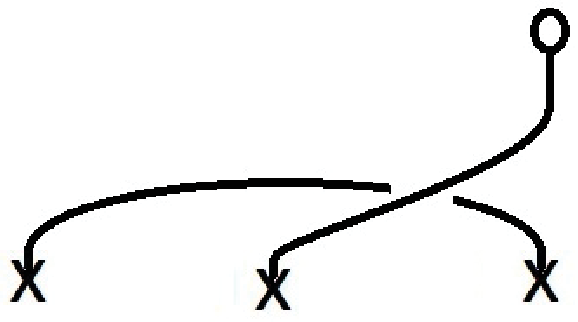}\end{gathered}\cr
&=&\,\,\begin{gathered}\includegraphics[scale=0.4]{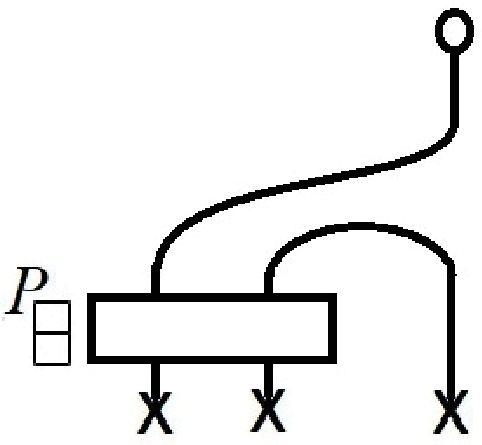}\end{gathered}
\label{fig:DiagMadag}
\eea

Using the definitions from above,
\bea
a^\dagger_i = a^\dagger\otimes e_i \in V_+\otimes W\qquad e_i \in W
\eea
There is an so($d$) equivariant map $\rho$
\bea
\rho : W \to (V_+\otimes W)
\eea
We can think of this as the map which attaches $e_i \in W$ to $a^\dagger$ to produce $a^\dagger_i = a^\dagger\otimes e_i$.
The map commutes with so($d$). This leads to the definition of a vector space of diagrams
\bea
V^*=\oplus_{n=0}^\infty {\rm Hom}_{so(d):d\gg n}(W_2^{\otimes n}\otimes W,V_+\otimes W)
\eea

We can define a diagrammatic inner product for $V^*$ (which involves loops evaluating to $d$) and show that 
$V^*\otimes V^*$ contains orthogonal subspaces corresponding to the symmetric-traceless, the trace, and the anti-symmetric.
The proof proceeds by proving $B_{n=2}(d)$ commutes with $U^*$ action on $V^*\otimes V^*$.
Much as $B_2(d)$ commutes with Uso($d$) on $V_d\otimes V_d$, but in the above both the algebra and the representation
space are spanned by diagrams. 
$d$ appears upon evaluation of Casimirs, which involve loops evaluated as $d$ - which can then be set to $4-\epsilon$.

There are some rather natural conjectures we can formulate about $(V^{*})^{\otimes n}$. 
First, the action of $U^*$ should commute with a known diagrammatic algebra, the Brauer algebra $B_d(n)$, much as $B_d(n)$ commutes with $Uso(d)$ in $V^{\otimes n}_d$. 
Proving this conjecture would involve generalising arguments given in \cite{deMelloKoch:2020roo}.
These are the first steps towards a fully diagrammatic Schur-Weyl duality where $U^*$, with loop parameter $d$, acts on
$V^{*\otimes n}$ and is Schur-Weyl dual to $B_d(n)$. 

\section{Summary and Outlook}

Our key result \cite{deMelloKoch:2020roo} has been to define $U_{*,2}$ acting on $V^{*,2}$ as a generic $d$ 
version of Uso($d$,2) acting on $\tilde{V}$. 
To summarise, we present evidence that perturbative CFT can be formulated in terms of Uso($d$,2) (for theories 
in integer dimensions) or $U_{*,2}$ (for theories like Wilson Fischer), using familiar constructions in algebra/representation
theory, namely
\begin{itemize}
\item indecomposable representations,
\item deformed co-products and
\item diagram algebras.
\end{itemize}
\begin{acknowledgement}
SR is supported by the STFC consolidated grant ST/P000754/1 `` String Theory, Gauge Theory \& Duality” and  a Visiting Professorship at the University of the Witwatersrand. RdMK is supported by a Simons Foundation Grant Award ID 509116 and by the South African Research Chairs initiative  of the Department of Science and Technology and the National Research Foundation. We thank Matt Buican and Adrian Padellaro for useful discussions  on the subject of this paper.  
\end{acknowledgement}


\begin{thebibliography}{99.}%
%
%


%
\bibitem{Maldacena:1997re}
J.~M.~Maldacena,
``The Large N limit of superconformal field theories and supergravity,''
Adv. Theor. Math. Phys. \textbf{2}, 231-252 (1998)
doi:10.1023/A:1026654312961
[arXiv:hep-th/9711200 [hep-th]].

\bibitem{Argyres:1995jj}
P.~C.~Argyres and M.~R.~Douglas,
``New phenomena in SU(3) supersymmetric gauge theory,''
Nucl. Phys. B \textbf{448}, 93-126 (1995)
doi:10.1016/0550-3213(95)00281-V
[arXiv:hep-th/9505062 [hep-th]].

\bibitem{Seiberg:1997zk}
N.~Seiberg,
``New theories in six-dimensions and matrix description of M theory on T**5 and T**5 / Z(2),''
Phys. Lett. B \textbf{408}, 98-104 (1997)
doi:10.1016/S0370-2693(97)00805-8
[arXiv:hep-th/9705221 [hep-th]].

\bibitem{Rattazzi:2008pe}
R.~Rattazzi, V.~S.~Rychkov, E.~Tonni and A.~Vichi,
``Bounding scalar operator dimensions in 4D CFT,''
JHEP \textbf{12}, 031 (2008)
doi:10.1088/1126-6708/2008/12/031
[arXiv:0807.0004 [hep-th]].

\bibitem{FMS}  
P. Di Francesco, P. Mathieu, D. Sénéchal, ``Conformal Field Theory,'' Graduate Texts in Contemporary Physics, Springer 1997

\bibitem{VOM} 
I Frenkel, J. Lepowsky, A. Meurman ``Vertex operator algebras and the Monster,'' Volume 134, Pure and Applied Mathematics, 1989 

\bibitem{Madore}
J.~Madore,
``The Fuzzy sphere,''
Class. Quant. Grav. \textbf{9} (1992), 69-88
doi:10.1088/0264-9381/9/1/008

\bibitem{Podles} 
P. Podles, ``Quantum spheres,'' Lett. Math. Phys. 14, 193-202 (1987)

\bibitem{NM}  
M. Noumi, K. Mimachi ``Quantum 2-Spheres and Big q-Jacobi Polynomials,'' Commun. Math. Phys. 128, 521-531 (1990)

\bibitem{Corley:2001zk}
S.~Corley, A.~Jevicki and S.~Ramgoolam,
``Exact correlators of giant gravitons from dual N=4 SYM theory,''
Adv. Theor. Math. Phys. \textbf{5}, 809-839 (2002)
doi:10.4310/ATMP.2001.v5.n4.a6
[arXiv:hep-th/0111222 [hep-th]].

\bibitem{Balasubramanian:2004nb}
V.~Balasubramanian, D.~Berenstein, B.~Feng and M.~x.~Huang,
``D-branes in Yang-Mills theory and emergent gauge symmetry,''
JHEP \textbf{03}, 006 (2005)
doi:10.1088/1126-6708/2005/03/006
[arXiv:hep-th/0411205 [hep-th]].

\bibitem{deMelloKoch:2007rqf}
R.~de Mello Koch, J.~Smolic and M.~Smolic,
``Giant Gravitons - with Strings Attached (I),''
JHEP \textbf{06}, 074 (2007)
doi:10.1088/1126-6708/2007/06/074
[arXiv:hep-th/0701066 [hep-th]].

\bibitem{Brown:2007xh}
T.~W.~Brown, P.~J.~Heslop and S.~Ramgoolam,
``Diagonal multi-matrix correlators and BPS operators in N=4 SYM,''
JHEP \textbf{02}, 030 (2008)
doi:10.1088/1126-6708/2008/02/030
[arXiv:0711.0176 [hep-th]].

\bibitem{Bhattacharyya:2008rb}
R.~Bhattacharyya, S.~Collins and R.~de Mello Koch,
``Exact Multi-Matrix Correlators,''
JHEP \textbf{03}, 044 (2008)
doi:10.1088/1126-6708/2008/03/044
[arXiv:0801.2061 [hep-th]].

\bibitem{Mattioli:2014yva}
P.~Mattioli and S.~Ramgoolam,
``Quivers, Words and Fundamentals,''
JHEP \textbf{03}, 105 (2015)
doi:10.1007/JHEP03(2015)105
[arXiv:1412.5991 [hep-th]].

\bibitem{Kimura:2014mka}
Y.~Kimura,
``Multi-matrix models and Noncommutative Frobenius algebras obtained from symmetric groups and Brauer algebras,''
Commun. Math. Phys. \textbf{337}, no.1, 1-40 (2015)
doi:10.1007/s00220-014-2231-6
[arXiv:1403.6572 [hep-th]].

\bibitem{Ramgoolam:2016ciq}
S.~Ramgoolam,
``Permutations and the combinatorics of gauge invariants for general N,''
PoS \textbf{CORFU2015}, 107 (2016)
doi:10.22323/1.263.0107
[arXiv:1605.00843 [hep-th]].

\bibitem{deMelloKoch:2014aot}
R.~de Mello Koch and S.~Ramgoolam,
``CFT$_4$ as $SO(4,2)$-invariant TFT$_2$,''
Nucl. Phys. B \textbf{890}, 302-349 (2014)
doi:10.1016/j.nuclphysb.2014.11.013
[arXiv:1403.6646 [hep-th]].

\bibitem{deMelloKoch:2018klm}
R.~de Mello Koch and S.~Ramgoolam,
``Free field primaries in general dimensions: Counting and construction with rings and modules,''
JHEP \textbf{08}, 088 (2018)
doi:10.1007/JHEP08(2018)088
[arXiv:1806.01085 [hep-th]].

\bibitem{deMelloKoch:2020roo}
R.~de Mello Koch and S.~Ramgoolam,
``Perturbative 4D conformal field theories and representation theory of diagram algebras,''
JHEP \textbf{05}, 020 (2020)
doi:10.1007/JHEP05(2020)020
[arXiv:2003.08173 [hep-th]].

\bibitem{Atiyah} 
M. F. Atiyah, ``Topological quantum field theory,'' Publications Mathématiques de l'IHÉS, 1988
 
\bibitem{Kock} 
J. Kock, ``Frobenius Algebras and 2-D Topological Quantum Field Theories,'' CUP 2010

\bibitem{Binder:2019zqc}
D.~J.~Binder and S.~Rychkov,
``Deligne Categories in Lattice Models and Quantum Field Theory, or Making Sense of $O(N)$ Symmetry with Non-integer $N$,''
JHEP \textbf{04}, 117 (2020)
doi:10.1007/JHEP04(2020)117
[arXiv:1911.07895 [hep-th]].

\bibitem{Moore:2006dw}
G.~W.~Moore and G.~Segal,
``D-branes and K-theory in 2D topological field theory,''
[arXiv:hep-th/0609042 [hep-th]].

\bibitem{Liendo:2017wsn}
P.~Liendo,
``Revisiting the dilatation operator of the Wilson\textendash{}Fisher fixed point,''
Nucl. Phys. B \textbf{920}, 368-384 (2017)
doi:10.1016/j.nuclphysb.2017.04.020
[arXiv:1701.04830 [hep-th]].

\bibitem{Beisert:2003tq}
N.~Beisert, C.~Kristjansen and M.~Staudacher,
``The Dilatation operator of conformal N=4 superYang-Mills theory,''
Nucl. Phys. B \textbf{664}, 131-184 (2003)
doi:10.1016/S0550-3213(03)00406-1
[arXiv:hep-th/0303060 [hep-th]].
\end{thebibliography}
\end{document}